\documentclass[12pt,a4paper]{article}

\usepackage[nosort]{cite}
\usepackage[hyperref,bulletsep]{collect}
\usepackage[active]{srcltx}
\usepackage{graphicx}
\usepackage{pst-all}
\usepackage{bbm}
\usepackage{amsmath}
\usepackage{amssymb}
\usepackage{subfigure}
\usepackage{array}

\renewcommand{\includegraphics}[1]{}

\def\IR{\mathbb{R}}

\def\IZ{\mathbb{Z}}
\def\IP{\mathbb{P}}

\def\id{\protect{{1 \kern-.28em {\rm l}}}}

\def\be{\begin{eqnarray}}
\def\ee{\end{eqnarray}}

\def\nn{\nonumber}

\makeatletter
\renewcommand\section{\@startsection {section}{1}{\z@}%
                                   {-3.5ex \@plus -1ex \@minus -.2ex}%
                                   {2.3ex \@plus.2ex}%
                                   {\normalfont\large\bfseries}}
\renewcommand\subsection{\@startsection{subsection}{2}{\z@}%
                                   {-3.25ex\@plus -1ex \@minus -.2ex}%
                                   {1.5ex \@plus .2ex}%
                                   {\normalfont\normalsize\bfseries}}
\makeatother

\def\bfsigma{{\boldsymbol{\sigma}}}
\def\AdSS{AdS$_5\times$S$^5$ }
\def\AdSP{AdS$_4\times\IP^3$ }

\begin{document}

\thispagestyle{empty}
\begin{flushright}\footnotesize
\texttt{AEI-2008-051}
\vspace{0.8cm}
\end{flushright}

\renewcommand{\thefootnote}{\fnsymbol{footnote}}
\setcounter{footnote}{0}

\begin{center}
{\Large\textbf{\mathversion{bold}
Spinning strings at one-loop in AdS$_4\times \IP^3$
}\par}

\vspace{1.5cm}

\textrm{Tristan McLoughlin$^1$, Radu Roiban$^2$} \vspace{8mm}

\textit{$^{1}$
Max-Planck-Institut f\"ur Gravitationsphysik\\
Albert-Einstein-Institut\\
Am M\"uhlenberg 1, 14476 Potsdam, Germany}\\
\vspace{3mm}

\textit{$^{2}$
Department of Physics, Pennsylvania State University\\
University Park, PA 16802, USA}\\
 \vspace{5mm}
 
\textit{$^{1}$}\texttt{tristan.mcloughlin@aei.mpg.de},\ \textit{$^{2}$}\texttt{radu@phys.psu.edu}  

\par\vspace{1cm}

\textbf{Abstract} \vspace{5mm}

\begin{minipage}{14cm}

We analyze the folded spinning string in AdS$_4\times \IP^3$ with spin
$S$ in AdS$_4$ and angular momentum $J$ in $\IP^3$. We calculate the
one-loop correction to its energy in the scaling limit of both $\ln S$
and $J$ large with their ratio kept fixed. This result should
correspond to the first subleading strong coupling correction to the
anomalous dimension of operators of the type $Tr(D^S
(Y^{\dagger}Y)^J)$ in the dual ${\cal N}=6$ Chern-Simons-matter
theory.
Our result appears to depart from the predictions for the generalized
scaling function found from the all-loop Bethe equations conjectured
for this AdS$_4$/CFT$_3$ duality. We comment on the possible origin of
this difference.

\end{minipage}

\end{center}

\vspace{0.5cm}

\newpage
\setcounter{page}{1}
\renewcommand{\thefootnote}{\arabic{footnote}}
\setcounter{footnote}{0}

\tableofcontents


\newpage
\section{Introduction}

The spinning folded string in AdS$_5$ has played an important role in
our quantitative understanding of the AdS/CFT duality. In the large
spin limit, the difference between its energy $E$ and spin $S$
scales like $\ln S$ \cite{Gubser:2002tv}; the proportionality
coefficient is the universal scaling function $f(\lambda)$ which
provided the first controlled example of an interpolating function
between weak and strong coupling.  These spin $S$ states are thought
to be dual to the operator tr$(ZD^SZ)$ where $D$ is the light-cone
covariant derivative and $Z$ is one of the complex scalar fields of
the theory; for such operators the logarithmic scaling has long been
known
\cite{Korchemsky:1988si,Korchemsky:1992xv,Bassetto:1993xd}.

A spinning folded string also exists in sigma models on
lower-dimensional AdS spaces, such as AdS$_4\times\IP^3$; it 
was pointed out in  \cite{Aharony:2008ug} that in the large spin limit
they have similar properties as the AdS$_5$ state, that is
\be
E-S\propto \ln S+{\cal O}(S^0)~~.
\ee
The gauge theory dual to closed string theory on AdS$_4\times\IP^3$
was recently conjectured to be a certain ${\cal N}=6$ superconformal
three-dimensional Chern-Simons theory \cite{Aharony:2008ug} (see also 
\cite{Benna:2008zy}). 
At finite $N$ and $k$, this $U(N)\times U(N)$ gauge theory is in fact
thought to describe the low-energy physics of $N$ M2-branes on
$\IR^{1,2}\times \mathbb{C}^4/{\IZ}_k$, where $k$ is interpreted as the
level of the Chern-Simons theory (for recent discussions on the
M2-brane worldvolume theory see e.g.
\cite{Schwarz:2004yj,Bagger:2007vi,Bagger:2007jr,Bagger:2006sk,
Gustavsson:2007vu,Gustavsson:2008dy,Distler:2008mk,Lambert:2008et}); 
in the large $N$ limit the gravity dual becomes M-theory on
AdS$_4\times $S$^7/{\IZ}_k$ where the orbifold group lies inside a
$U(1)$ subgroup of the $SO(8)$ isometry group of $S^7$. This theory
also has an 't Hooft limit where both $k$ and $N$ are taken to be
large with $\lambda =N/k$ kept fixed. In this limit the size of the
circle fiber acted upon by the ${\IZ}_k$ orbifold becomes very small
and thus the appropriate description is as type IIA theory on
AdS$_4\times\IP^3$.
The ${\cal N}=6$ Chern-Simons theory \cite{Aharony:2008ug} exhibits an
$SU(4)\times U(1)$ global symmetry group, the first factor of which is
the R-symmetry. In addition to the gauge-fields, it also contains
eight bi-fundamental scalar fields $Y^I$ and $Y_I^{\dagger}$ which
transform as ${\mathbf 4}_{\mathbf +1}$ and ${\mathbf {\bar
4}}_{\mathbf -1}$ of $SU(4)\times U(1)$. 
The representations of the eight fermionic bi-fundamental
superpartners follow from the representation of the supercharges; for
the M2-brane theory the supercharges transformed as the ${\mathbf
8}_c$ representation of the $SO(8)$ R-symmetry and decompose under the
commutant of the orbifold action as ${\mathbf 6}_{\mathbf 0}\oplus
{\mathbf 1}_{\mathbf 2}\oplus {\mathbf 1}_{\mathbf -2}$. It is natural
to expect that the spinning folded strings should be dual to single
trace gauge invariant operators made of a large number of covariant
derivatives and some finite number of other fields.

The twist-two operators tr$(ZD^SZ)$ of ${\cal N}=4$ SYM theory are not
captured by the asymptotic Bethe ansatz. To bypass this problem and,
at the same time, to make a cleaner identification between the gauge theory
operators and string solutions it is useful to generalize the rotating
folded string by adding a further angular momentum $J$ in the compact
space. The dual operators tr$(D^SZ^J)$ belong to the $sl(2)$ sector of the
theory. For strings in \AdSS   this has been done in 
\cite{Frolov:2002av}. The resulting target space energy,
$E(\sqrt{\lambda}, S, J)$, is a nontrivial function of its arguments
and may be expanded in different regimes, uncovering and testing
various aspects of the gauge and string Bethe ans\"atze. 
One can straightforwardly find similar strings moving along an
$S^1\subset \IP^3$ with angular momentum $J$. Invariance under
$U(N)\times U(N)$ gauge transformations, and the requirement that the
operator be charged only under one Cartan generator of the R-symmetry
group suggests that the relevant operators are 
tr$(D^S (Y^1Y_4^{\dagger})^J)$.
\footnote{Here we assigned charges to the
fields in the ${\mathbf 4}$ of $SU(4)$ such that $Y^4$ has equal
charges under all three Cartan generators while $Y^i$ with $i=1,2,3$
has the same charge as $(Y_4)^\dagger$ under the $i$-th generator and
the charge as $Y^4$ under the other two generators.}
\footnote{
It is worth noting that two scalar fields together with a covariant
derivative can carry the same quantum numbers as a fermion bilinear so
that generically such states will mix; with a some care however, it
is still possible to identify a closed sl$(2)$ sector. }

As for the ${\cal N}=4$ theory, the dilatation operator of the
Chern-Simons theory appears to be described by an integrable spin
chain at weak coupling \cite{Minahan:2008hf} (see also
\cite{Gaiotto:2008cg,Bak:2008cp}). Unlike that of the ${\cal N}=4$
theory this spin chain is alternating due to the presence of fields
in the bifundamental representation.
Given as $J$ roughly corresponds to the spin-chain length, it is
necessary to take it to be large in order to expect an exact Bethe ansatz,
which would therefore be asymptotic.
The choice of vacuum for the spin chain leaves unbroken a symmetry
group similar to that of the spin chain of ${\cal N}=4$ SYM
theory. Together with information \cite{Gromov:2008bz} extracted from
a conjectured worldsheet action for strings in \AdSP 
\cite{Arutyunov:2008if,Stefanski:2008ik,Fre:2008qc}, asymptotic Bethe 
equations have been conjectured in \cite{Gromov:2008qe} (see also
\cite{Ahn:2008aa}). To leading order in the weak coupling expansion
these equations reproduce the results of direct anomalous dimension
calculations \cite{Minahan:2008hf}.  
Similarly to \AdSS, the study of the properties of classical string solutions, 
such as the finite size
corrections to their energy, (see
\cite{Ahn:2008hj,Lee:2008ui,Astolfi:2008ji,Chen:2008qq,Ahn:2008gd,
Grignani:2008te}) may be used to carry out further tests of the 
Bethe equations.

In this work we will consider the one-loop string corrections 
to the energy of the spinning folded string in \AdSP.
 While the full superstring action on this space is not known, sigma
 models based on the coset $OSp(6|4)/SU(3)\times U(1)\times SO(3,1)$
 and supplied with an appropriate Wess-Zumino like term
 \cite{Arutyunov:2008if,Stefanski:2008ik,Fre:2008qc} have been
 suggested to represent partially $\kappa$-gauge fixed Green-Schwarz
 string actions. Furthermore, it has been shown that these actions are classically
 integrable suggesting that it may be possible to study this theory using similar
 methods  to the \AdSS case. 
We will however not use these actions. To one-loop order only the
quadratic part of the fermion action is necessary and its structure is
well-known in terms of the supersymmetric covariant derivative. 

After recalling the supergravity background \cite{Aharony:2008ug} in
\S2 we proceed in \S3 to discuss the spinning string solutions in
\AdSP, some of their scaling limits as well as the
expectations for the semiclassical expansions of their energy, all of which
are quite analogous to those of spinning strings in \AdSS.
In  \S4 we find the spectrum of bosonic and fermionic
fluctuations  around the spinning folded string solution in the
scaling limit. In \S5 we evaluate the one-loop correction to the
target space energy  both for strings with $J=0$ and $J\neq0$ in the semi-classical 
scaling limit. We show that the
quadratic and logarithmic divergences cancel and extract the one-loop 
correction to the generalized scaling function. 
In \S6 we discuss the comparison with the Bethe ansatz predictions and discuss some
possible future directions.

\section{AdS$_4 \times \IP^3$ Background}
\label{sec:Background}
Recently, \cite{Aharony:2008ug}, it was pointed out that the near
horizon geometry of M2-branes on a special $\mathbb{Z}_k$ quotient of flat space
is, for large values of $k$, AdS$_4\times \IP^3$. Taking the standard
M2-brane near horizon geometry of AdS$_4\times $S$^7$ and writing the
S$^7$ as a S$^1$ fibration over $\IP^3$ the effect of the
$\mathbb{Z}_k$ quotient is simply to make the radius of the S$^1$
smaller by a factor of $k$. The compactification from eleven to ten
dimensions gives rise to a two form flux which is proportional to the
K\"ahler form on the $\IP^3$ and the four form flux is unaffected
except that the number of units of flux is reduced by a factor of
$k$. To be more explicit the background fields after the quotient are
\be
ds^2&=&\frac{R^3}{4k}\left(ds_{AdS_4}^2+4ds^2_{\IP^{3}}\right)
~~~~~~~~~~
e^{2\phi}=\frac{R^3}{k^{3}}
\cr
F_{2}&=&k\ J_{\IP^{3}}
~~~~~~~~~~~~~~
F_{4}=\frac{3}{8}R^3{\rm Vol}_{AdS_4} 
\label{bkgrnd}
\ee
Above, the metric and the forms are written in terms of those of
spaces of unit radius. For AdS$_4$ we use global coordinates,
$\left(t,\rho,\theta,\phi\right) $ and the resulting metric is the
standard
\be
\label{ads_metric}
ds^2_{{\rm AdS}_4}&=&
-\cosh^2\rho\ dt^2+d\rho^2+\sinh^2\rho 
\left(d\theta^2+\sin^2\theta d\phi^2\right)
\ee
and we make use of the parameterization, \cite{Hoxha:2000jf}, of the $\IP^3$
geometry in terms of the coordinates $\left(\zeta_1,\zeta_2,\zeta_3,\tau_1,\tau_2,\tau_3\right)$,
\be
\label{p3_metric}
ds^2_{\IP^{3}}&=& d\zeta_1^2
+\sin^2\zeta_1\left[ d\zeta_2^2+\cos^2\zeta_1\left(d\tau_1+\sin^2\zeta_2\left(d\tau_2+\sin^2\zeta_3d\tau_3\right)\right)^2\right. \nn\\
& &\left. + \sin^2\zeta_2\left( d\zeta_3^2
+\cos^2\zeta_2\left(d\tau_2+\sin^2\zeta_3d\tau_3\right)^2+\sin^2\zeta_3\cos^2 \zeta_3 d\tau_3^2 \right)\right]
\ee
where we have pulled out an overall factor of $R^2_{\rm AdS}=R^3/4k$
with $R$ being the radius of the original AdS$_4\times $S$^7$
geometry. This expression for the $\IP^3$ metric can be found by
iteratively embedding $\IP^{n-1}$ in $\IP^n$. The two-form can be
written as the exterior derivative, $F_{2} =k d\omega$, of a one-form
defined locally by
\be
\omega= \sin^2 \zeta_1
\left(d\tau_1+\sin^2\zeta_2\left(d\tau_2+\sin^2\zeta_3 
d\tau_3\right)\right).
\ee
 In physical coordinates one has:
\be
(F_2)_{\mu\nu}=2\frac{k^2}{R^3}J_{\mu\nu}
~~~~~~~~
(F_4)_{abcd}={6}\frac{k^2}{R^3}\epsilon_{abcd}
\ee
or 
\be
e^{\phi}(F_2)_{\mu\nu}=\frac{1}{R_{\rm AdS}}J_{\mu\nu}
~~~~~~~~
e^{\phi}(F_4)_{abcd}=\frac{3}{R_{\rm AdS}}\epsilon_{abcd}
\ee
where $J$ and $\epsilon$ are numerical tensors with entries $\pm1$ and
$0$. They are, respectively, the entries of the K\"ahler form and of the
volume form on unit $\IP^3$ and AdS$_4$. Finally the ten-dimensional
radius of curvature will be related to the 't Hooft coupling by
\be
R^2_{\rm string}=\frac{R^3}{k}=2^{5/2}\pi \sqrt{\lambda}.
\ee
We now turn to the study of a particular class of spinning strings in
this background.

\section{Spinning string solution and scaling limits}
\label{sec:Spinning_string}

Many of the spinning string solutions of Frolov and Tseytlin
\cite{Frolov:2002av, Frolov:2003qc, Frolov:2003tu} are again 
solutions of strings on AdS$_4\times \IP^3$ and indeed many of their
calculations, including that of the quantum correction to the long
spinning string, are modified only very slightly. Let us briefly
summarize some of the relevant details about spinning strings. We wish
to consider folded closed strings that have two non-vanishing charges: 
one spin, $S$, in the AdS$_4$ space and one angular momentum, $J$, 
in the compact $\IP^3$ and that are solutions of the equations of motion of the action
\be
I&=&I_{\rm AdS_4}+I_{\IP^3}\nn\\
 &=&\frac{R^2_{\rm AdS}}{4 \pi}\int \ d\tau d\sigma\ \sqrt{h}h^{ab } 
\left(G_{\mu\nu}^{\rm AdS}\partial_a X^{\mu} \partial_b X^{\nu}
   +4 G_{\mu\nu}^{\IP^3}\partial_a X^{\mu} \partial_b
   X^{\nu}\right)~~.  
\label{bose_action}
\ee 
Due to the choice of spins, the solution fits inside an 
AdS$_3\times$S$^1$ subspace and it is in fact identical to that of GKP,
\cite{Gubser:2002tv} and further studied in
\cite{Frolov:2002av},  except for a multiplication of the S$^1$ angular
momentum parameter by $\frac{1}{2}$. This is a consequence of the
numerical factor in the second term in the action (\ref{bose_action}).

As in \AdSS, the worldsheet semiclassical expansion about these spinning
string solutions is naturally organised as an expansion in
$\tfrac{1}{\sqrt{2\lambda}}$ (which is proportional to the inverse
string tension) which keeps fixed the charge densities ${\cal
S}=\frac{S}{\sqrt{2\lambda}}$ and ${\cal
J}=\frac{J}{\sqrt{2\lambda}}$.  The target space energy of the string
is given by
\be
E=\sqrt{2\lambda}\ {\cal E}\left({\cal S},{\cal J}, 
\frac{1}{\sqrt {2\lambda}}\right)
=\sqrt{2\lambda}\left[{\cal E}_0\left({\cal S},{\cal J}\right)
+\frac{1}{\sqrt{2\lambda}}{\cal E}_1\left({\cal S},{\cal J}\right)
+\dots \right].
\ee

Given the complexity of the solution \cite{Frolov:2002av} additional
limits are useful. 
We will consider  the so-called ``semi-classical scaling"
or long-string limit of the spinning string solutions, see 
\cite{Frolov:2002av, Frolov:2006qe} and also
\cite{Roiban:2007ju},
\be
\label{eqn:scaling_limit}
{\cal S}\gg {\cal J}\gg1, \qquad {\rm with}\ \ell \equiv 
\frac{\cal J}{2\ {\ln{\cal S}}}\  {\rm fixed}.
\ee
Since we are interested in the limits $\ln {\cal S}\gg \ln {\cal J}$ and ${\cal
S}\gg \sqrt{2\lambda}$ this equivalent to
\be
{S}\gg { J}\gg1, \qquad {\rm with}\  \ell\approx \frac{J}{2\sqrt{2\lambda}\ 
{\ln{ S}}} \ {\rm fixed}.
\ee

As discussed at length in \cite{Frolov:2006qe,Roiban:2007ju} , in this
limit the solution simplifies dramatically becoming
homogeneous. Choosing \footnote{ There are many different $S^1$
factors that one may pick inside $\IP^3$. A particularly useful
choice, which leads to the vanishing of some components of the spin
connection, may be identified by introducing new coordinates
\be
\tau_1=\varphi_3-\beta,\qquad \tau_2=\beta-\gamma,\qquad \tau_3=\varphi_3+\gamma~~.
\ee

with all the other coordinates set to zero. } the angle $\varphi_3$
parametrizing the circle S$^1\subset \IP^3$ as
$\varphi_3=\tfrac{1}{2}(\tau_1+\tau_2+\tau_3)$, the relevant part of
the action is given by the metric
\be
ds^2=R^2_{\rm AdS}\left(d\rho^2-\cosh^2\rho\ dt^2+\sinh^2\rho\ d\phi^2
+ 4d\varphi_3^2\right)~~.
\ee
Then, the solution is just
\be
{\bar t}=\kappa\tau~~~~~~~~
{\bar \phi}=\kappa\tau~~~~~~~~
{\bar \rho}=\mu \sigma~~~~~~~
{\bar \varphi}_3=\frac{1}{2}\nu\tau~~~~~~~\mu =\sqrt{\kappa^2-\nu^2}~~;
\label{sol_0}
\label{eqn:scaling_solution}
\ee
the other \AdSP  coordinates take constant values, the
nonvanishing ones being
\be
{\bar  \theta}=\frac{\pi}{2},\qquad
{\bar \zeta}_1=\frac{\pi}{4},\qquad
 {\bar \zeta}_2=\frac{\pi}{2},\qquad
  {\bar \zeta}_3=\frac{\pi}{2}~~.
\ee
As with all classical solutions, two-dimensional Lorentz invariance is
spontaneously broken. As we shall see it turns out to be convenient to express the
solution in terms of constant vectors. In this way, Lorentz
invariance is apparently preserved (and it would be if one allowed these
constant vectors to transform as implied by the indices they carry). In
analogy with the spinning string solution in AdS$_5\times$S$^5$, we
define the vectors ${\hat {\rm n}}$, ${\tilde {\rm n}}$ and ${\hat {\rm m}}$
\be
d{\bar t}={\hat {\rm n}}\cdot d\bfsigma~~~~~~
d{\bar \phi}={\hat {\rm n}}\cdot d\bfsigma~~~~~~
d{\bar \rho}={\tilde {\rm n}}\cdot d\bfsigma~~~~~~
d{\bar \varphi}_3=\frac{1}{2}{\hat {\rm m}}\cdot d\bfsigma~~~~~~
\bfsigma=(\sigma^0,\sigma^1)\equiv(\tau,\sigma)~~.
\label{vectors}
\ee
The Virasoro constraint relates these vectors as follows:
\be
\eta^{ab}{\hat {\rm n}}_a{\hat {\rm n}}_b+\eta^{ab}{\tilde {\rm
n}}_a{\tilde {\rm n}}_b
=\eta^{ab}{\hat {\rm m}}_a{\hat {\rm m}}_b=-\nu^2~.
\ee

We must also impose periodicity in the $\sigma$ direction, ${\bar
\rho}(\sigma+2\pi)={\bar \rho}(\sigma)$, which is satisfied by
 interpreting the solution (\ref{sol_0}) as a string folded
onto itself. The string is thus made of four segments: for, $0\leq
\sigma\leq \tfrac{\pi}{2}$, ${\bar \rho}$ increases from $0$ to its
maximum $\rho_0$, while for $\tfrac{\pi}{2}\leq \sigma\leq \pi$ it
decreases from its maximum value back to zero and then repeats. The
relation between the parameters of the solution, $\kappa,\,\mu $ and
$\nu$, is a consequence of the Virasoro constraint. We note that for
the above solution, being in the scaling limit
\eqref{eqn:scaling_limit}, $\kappa$ and $\mu $ are both large while
$\ell=\tfrac{\nu}{\mu }$ is kept fixed. This can be seen clearly by
relating the parameters of the solution to the global charges of the
string which are given by, $E=
\sqrt{2\lambda}\ {\cal E}$ etc, with
%
\be 
{\cal E}=\int^{2\pi}_0 d\sigma\ \frac{\kappa}{2} \cosh^2 {\bar
\rho},& &\qquad {\cal S}=\int^{2\pi}_{0} d\sigma\ \frac{\kappa}{2}
\sinh^2 {\bar \rho},\nn\\ & &\kern-50pt {\cal J}=\int^{2\pi}_{0}
d\sigma\ \nu.  
\label{charges}
\ee

We thus have
\be 
\mu =\frac{1}{\pi}\ln{\cal S}, \qquad \mu \gg
 1,\qquad \ell=\frac{\nu}{\mu }={\rm fixed}.  
\ee 
If we rescale $\sigma$ by $\mu $ we get $\rho=\sigma$ but now the
worldsheet has length $L=2\pi \mu \sim \ln S\gg 1$ and in the strict
$L\rightarrow \infty$ limit the closed string can be thought of as two
infinite overlapping open strings. In this limit we can neglect all
effects of the string end points where from the closed string point of
view the worldsheet curvature becomes infinite. In the scaling limit
\eqref{eqn:scaling_limit} we have ${\cal E}={\cal S}+\kappa \ \pi$ and
thus to leading order
\be
{\cal E}_0-{\cal S}=\mu  \pi\sqrt{1+\ell^2} =
\ln {\cal S}\sqrt{1+\frac{{\cal J}^2}{4 \ \ln^2 {\cal S}}}
\ee
or using the fact that ${\cal S}\gg {\cal J}$ and $\tfrac{\cal S}{\sqrt{2\lambda}}\gg1$
\be
E_0-S&=&\sqrt{2 \lambda}\ln S\sqrt{1+\frac{J^2}{8\lambda\ln^2 S}}\nn\\
&=&\sqrt{2\lambda} f_0(\ell)\ln S~.  
\label{cusp_CS_0}
\ee 
We can of course consider the limit in which the angular momentum in
the compact space is vanishing, or more precisely the limit
$\tfrac{\cal J}{\ln {\cal S}}\ll1$, the ``semi-classical scaling small"
limit. In this limit at leading order $E_0-S=\sqrt{2 \lambda} \ln S$
which is the result from
\cite{Aharony:2008ug}. 

Our aim here is to extend this result to include the next-to-leading
order correction to the spinning string energy which, as we shall
explicitly see, takes the form
\be
E_1={\rm f}_1(\ell)\ \ln S+\dots\ .
\ee
Thus, just as for the \AdSS string, it appears that the
strong coupling expansion in the scaling limit can be organised as
\be
E-S=\sqrt{2 \lambda}\  {\rm f}(\ell,\lambda)\ln S+\dots
\ee
and the function, ${\rm f}(\ell,\lambda)$ can be expanded in inverse powers of
$\sqrt{2\lambda}$ to give the coefficients ${\rm f}_0(\ell)$, ${\rm
f}_1(\ell)$, etc or alternatively one can first expand in powers of $\ell$
\be
{\rm f}(\ell,\lambda)={\rm f}(\lambda)+\ell^2 q(\ell, \lambda)
+\ell^4 p(\ell, \lambda)+\dots.
\ee
The function ${\rm f}(\lambda)$ is the three-dimensional analogue of
the universal scaling function $f(\lambda)$ of ${\cal N}=4$
super-Yang-Mills in four dimensions. Similarly to that case, we expect
that the functions $q(\ell, \lambda)$ and $p(\ell, \lambda)$ exhibit
logarithmic dependence on $\ell$ in the string coupling expansion.

It is perhaps worth mentioning that the relationship between GKP
spinning strings \cite{Gubser:2002tv} and the open strings dual to
light-like Wilson loops with a cusp, \cite{Kruczenski:2002fb}, that is
known to exist in AdS$_5$, persists in this context at least at the
level of the classical worldsheet.  The argument,
\cite{Kruczenski:2007cy}, that in the scaling limit, after an
analytic continuation combined with the use of the AdS isometries, these two string
solutions correspond to the same minimal surface is essentially
unchanged. 
Thus we expect the anomalous dimension of twist-two operators and the
cusp anomaly to be equal also in the dual three-dimensional
Chern-Simons theory. Their common value should define the scaling
function $f_{CS}(\lambda)$.\footnote{The coordinate transformations
relating the spinning folded string the the Wilson line with a cusp
can also be carried out in the presence of nonvanishing angular
momentum on S$^5$.}
This equivalence for the ${\cal N}=4$ theory was proven in weak
coupling perturbation theory
\cite{Korchemsky:1988si,Korchemsky:1992xv,Bassetto:1993xd} and
 in addition to the arguments cited above has been partially confirmed
 by direct calculation \cite{Kruczenski:2002fb,
 Frolov:2002av,Kruczenski:2007cy, Makeenko:2002qe}. It is also
 worthwhile mentioning that the same scaling function $f(\lambda)$
 governs the IR asymptotics of the gluon amplitude in the ${\cal N}=4$
 theory
 \cite{Sen:1981sd,Korchemsky:1985xj,Magnea:1990zb,Korchemsky:1993hr,
 Korchemskaya:1994qp,Sterman:2002qn,Bern:2005iz}. Furthermore for the
 four-point gluon amplitude it determines the finite
 part of the exponentiated all-loop expression found in
 \cite{Bern:2005iz,Anastasiou:2003kj}. In the context of the AdS/CFT
 correspondence the same functional dependence for the scattering
 amplitude was found at strong coupling by \cite{Alday:2007hr}. In
 large part this is entirely determined by the symmetries of the
 problem \cite{Drummond:2007aua,Alday:2007mf}. For AdS$_4$ we can, at
 least at strong coupling, formally find a similar relation though the
 interpretation, which makes use of several T-duality like
 transformations, is perhaps less clear.

\section{Fluctuation Spectrum}
\label{sec:fluctations}
\subsection{Bosonic action to quadratic order}
\label{subsec:Bosonic_fluctuations}

In this section we calculate the spectrum of bosonic quantum
fluctuations about the spinning string solutions, at least in the
homogeneous scaling limit. In this we will again follow very closely
\cite{Frolov:2002av, Frolov:2006qe} and so we will not belabor the
details - the calculations are essentially identical though with one
less transverse degree of freedom in the AdS space and one more in
the slightly more complicated $\IP^3$ space. The fluctuations about
the classical spinning string solution in the AdS$_4$ space are 
\be && t={\hat {\rm n}}\cdot\bfsigma
+\frac{{\tilde t}}{{\tilde \lambda}^{\frac{1}{4}}}~,\quad 
\rho= {\tilde {\rm n}}\cdot \bfsigma
+\frac{{\tilde \rho}}{{\tilde \lambda}^{\frac{1}{4}}}~,\quad 
\theta=\frac{\pi}{2}
+\frac{{\tilde \theta}}{{\tilde \lambda}^{\frac{1}{4}}}~,\quad  
\phi={\hat {\rm n}}\cdot\bfsigma
+\frac{{\tilde \phi}}{{\tilde \lambda}^{\frac{1}{4}}}~~.
\ee
In the above we have used as our expansion parameter \footnote{There
is some ambiguity in exactly what we use as the expansion parameter
however we fix this by demanding that for the analogous expansion
about the BMN string the energy of a single massive excitation is
$E-J=1+{\cal O}\left(\lambda\right)$.}  ${\tilde \lambda}={2 \pi^2
\lambda}$.  The bosonic action quadratic in fluctuations in the
AdS$_4$ space becomes 
\be
I_{{\rm AdS}_4}&=&-\frac{1}{4 \pi}\int d^2\sigma 
\Big[ (\partial \rho)^2-\cosh^2 {\bar \rho} (\partial t)^2+\left[
(\partial \phi)^2 +(\partial \theta)^2
-{\hat {\rm n}}\cdot {\hat {\rm n}} \,\theta^2\right] 
\sinh^2 {\bar \rho}\nn\\
& & \kern+80pt 
-2({\hat {\rm n}}\cdot \partial t-{\hat {\rm n}}\cdot\partial\theta)\kappa\ 
\rho\, \sinh(2{\bar \rho})\Big]
\ee
where we have dropped the tildes. To eliminate the explicit dependence
on ${\bar \rho}$ it is useful to redefine the fields as
\be
{\hat t}=\cosh {\bar \rho}\ t,\qquad {\hat \theta }=\sinh {\bar \rho}\
\theta,\qquad {\hat \phi}= \sinh {\bar \rho}\ \phi,\qquad {\hat \rho}=\rho
\ee
and do a further rotation in the $({\hat t},{\hat \phi})$ plane
\be
\chi={\hat \phi} \cosh {\bar \rho}- {\hat t}\sinh {\bar \rho},
\qquad \zeta =-{\hat \phi} \sinh {\bar \rho}+ {\hat t}\cosh {\bar \rho}~~
\ee
after which the action becomes 
\be
I_{{\rm AdS}_4}&=&-\frac{1}{4\pi}\int d^2\sigma \Big[ 
-(\partial { \zeta})^2+(\partial \chi )^2+(\partial {\hat \rho})^2
+4\zeta\,{\tilde {\rm n}}\cdot\partial\chi
+4{\hat \rho}\,{\hat {\rm n}}\cdot \partial\chi
+(\partial {\hat \theta})^2\nn\\
& &\kern+80pt +({\tilde {\rm n}}\cdot{\tilde {\rm n}}-{\hat {\rm n}}\cdot {\hat {\rm n}}){\hat \theta}^2
\Big]~~.
\ee

The spectrum is more conveniently expressed in terms of $\kappa$ and
$\nu$ rather than in terms of ${\hat {\rm n}}$ and ${\tilde {\rm n}}$.  Similarly to the
spectrum of bosonic fluctuations in AdS$_5\times $S$^5$, we find one
combination $\chi, \zeta$ and ${\hat \rho}$ being massless and one each
with dispersion relation
\be
\label{eqn:bosonic_ads_freq}
\omega_\pm(n)=\sqrt{n^2+2\kappa^2\pm 2\sqrt{\kappa^4+n^2\nu^2}},
\ee 
where here $n$ denotes the mode number. There is additionally one
transverse mode with mass squared $2\kappa^2-\nu^2$. For the string
moving on an $S^1$ inside the $\IP^3$ masses of the fluctuations are
quite straightforward with one longitudinal massless degree of
freedom, four with mass squared $\tfrac{\nu^2}{4}$ and one with mass
squared $\nu^2$. \footnote{It should be mentioned that, due to the
numerical factor in the second term on the right hand side of the
equation (\ref{bose_action}), the normalization of the quadratic term
of the $\IP^3$ fluctuations is non-standard. While this is irrelevant
at one-loop order, it must be carefully accounted for in higher-loop
calculations.} Note that in the absence of an angular momentum on
$\IP^3$, the spectrum exhibits the $SO(6)\simeq SU(4)$ symmetry of
$\IP^3$. For $J\propto\nu\ne 0$ this symmetry is broken to $SO(4)$.

As is the case for the \AdSS string, two of the massless
modes cancel against the contribution of the diffeomorphism ghosts
that arise from fixing conformal gauge. For a string spinning entirely
in AdS$_4$ we take $\nu$ to zero
and in this case the bosonic spectrum is particularly simple: we get
one massive excitation with $m^2=4\kappa^2$, one with $m^2=2\kappa^2 $ and
six massless modes so that $\sum_{\rm bosons}m^2=6\kappa^2$. As
discussed in \cite{Alday:2007mf} we can consider the fluctuations as
the Goldstone bosons (or fermions for the fermionic fluctuations to be
discussed in the next section). Thus we expect the six massless modes
from the $\IP^3$ to remain massless to all orders in worldsheet
perturbation theory.

\subsection{Fermionic action to quadratic order}
\label{subsec:Fermionic_fluctuations}

We now turn to the construction of the spectrum of fermionic
fluctuations.  As mentioned previously, the complete $\kappa$-gauge-invariant 
Green-Schwarz action on \AdSP is not known. Recently,
however, Green-Schwarz \cite{Arutyunov:2008if,Stefanski:2008ik} and
pure spinor \cite{Fre:2008qc} models based on the coset
$OSp(6|4)/SU(3)\times U(1)\times SO(1,3)$ have been constructed. 
The resulting sigma model possesses twenty-four fermionic degrees of
freedom and may be interpreted as a partial $\kappa$-gauge fixing of an
action with thirty-two fermionic degrees of freedom. The remaining
$\kappa$-symmetry generically removes eight of the fermions.
For strings moving entirely in AdS$_4$, such as the spinning folded
string, a larger number of degrees of freedom becomes unphysical; the
remaining $\kappa$-symmetry is enhanced and becomes capable of
removing twelve fermionic degrees of freedom, instead of eight 
\cite{Arutyunov:2008if}.

Such a small number of physical fermionic degrees of freedom
is not allowed by the usual rules for the Green-Schwarz string; 
 one would therefore expect that it is possible to use the supercoset
 models for the generalized spinning solutions with $J\neq0$ but not for
 $J=0$. 
Such a conclusion is, however, somewhat puzzling as we expect the
energy to be a smooth function of $J$. This motivates, in part, our
consideration of the generalized solutions where we can analyze the
$J\rightarrow0$ behavior and, separately, the $J=0$ solution.
\footnote{After this
work appeared a similar calculation using the coset approach
\cite{Alday:2008ut} was submitted which found that the $J\rightarrow 0$
limit was smooth and in agreement with our calculation.}

For our purposes we fortunately need only the the
quadratic-in-fermions part of the gauge-invariant Green-Schwarz action
and this is well known to have a standard expression in terms of the
target space covariant derivative:
\be
L_{2F}=i(\eta^{ab}\delta^{IJ}-\epsilon^{ab}s^{IJ}){\bar\theta}^Ie\llap/{}_a
D^{JK}_b\theta^K
\label{lag_0}
\ee
where $s={\rm diag}(1,-1)$ and $e_a^A$ is the pullback of the
vielbein
\be
e_a^A=\partial_a X^M E_M^A
\ee
(here $X^M$ denote generic target space coordinates). In type IIB
theory in the presence of a 5-form flux this expression was analyzed
in \cite{Frolov:2002av} and brought to a form resembling a
two-dimensional fermionic action.

We will analyze here the type IIA string theory, with additional
restrictions on the form of $D_a^{JK}$ due to the fluxes present in
the background (\ref{bkgrnd}). 
The structure of the action bears certain similarities with that in
the type IIB theory due to the fact that the background RR fluxes 
are constant on the tangent space.
Here however, the two fermions $\theta^1$ and
$\theta^2$ have opposite chiralities. Defining
$F\llap/{}_{(n)}=\frac{1}{n!}\Gamma^{N_1N_2\dots N_n}F_{N_1N_2\dots
N_n}$ the covariant derivative is
\be
D^{JK}_a&=&\left(\partial_a+\frac{1}{4}\partial_a X^M
\omega_M{}^{AB}\Gamma_{AB}\right)\delta^{JK} -\frac{1}{8}\partial_a
X^M E_M^A H_{ABC}\Gamma^{BC}(\sigma_3)^{JK}\cr
&+&\frac{1}{8}e^{\phi}\left[F_{(0)}(\sigma_1)^{JK}
                           +F\llap/{}_{(2)}(i\sigma_2)^{JK}
                    +F\llap/{}_{(4)}\!\!\!\!{}'~(\sigma_1)^{JK}\right] e\llap/{}_a
\ee
with $\sigma_i$ being the Pauli matrices and the modified form field
strength $F_4'$ given, as usual, by 
\be
F_4'=F_4-H\wedge C_1~~.
\ee

In the coordinates \eqref{ads_metric} the spin connection reads:
\be
\begin{array}{ll}
\omega^{01}=-\omega^{10}=\sinh\rho\; dt&~~
\omega^{21}=-\omega^{12}=\cosh\rho\; d\theta \\
\omega^{31}=-\omega^{13}=\cosh\rho\sin\theta \; d\phi&~~
\omega^{32}=-\omega^{23}=\cos\theta \; d\phi.
\end{array}
\ee

With regard to the spin connection for
the compact $\IP^3$ we note that using local Lorentz 
transformations it is always possible to choose
the spin connection to vanish along a chosen direction -- in
particular $\varphi_3$. It turns out that the coordinates
(\ref{p3_metric}) together with the choice of $\varphi_3$ mentioned above
realize this observation.
Thus, for spinning string solutions carrying a single charge in the
space transverse to AdS, the explicit form of the spin connection is
not necessary for the calculation of the spectrum of quadratic
fluctuations. If the profile in the transverse space involves a single
(isometric) field, then one also does not -- for the same purpose --
need to make sure that the full metric is written in the coordinates
adapted to the vanishing spin connection. Indeed, the spectrum is
invariant under coordinate transformations, so one can compute the
bosonic spectrum in any suitable coordinate system.

\subsubsection{The $(S,J=0)$ string}

Let us consider first the solution with vanishing angular momentum in
the transverse space. A reason for analyzing this configuration
separately (rather than as a limit of $J\ne 0$ configurations which
will be discussed later) is to test explicitly the continuity of the
energy and of the natural $\kappa$-gauge condition as a function of
$J$. Moreover, the details of the calculation compared to those for
the $J\ne 0$ configurations may point the origin of the enhancement of
the $\kappa$ symmetry of the $OSp(6|4)$ models.
As was exploited extensively in the calculation of one-loop
corrections to the energy of classical strings in \AdSS, no  bosonic
fluctuations appear in the quadratic fermion action; one simply
evaluates (\ref{lag_0}) on the classical solution. 
Using the 
fact that from (\ref{vectors}) we have ${\hat {\rm n}}=(\kappa, 0)$,
${\tilde {\rm n}}=(0,\kappa)$ and ${\hat {\rm m}}=(0,0)$ it follows that 
\be
e\llap/{}_a=\frac{R_{\rm string}}{2}
\left[{\hat {\rm n}}_a(\cosh{\bar \rho}\Gamma_0+\sinh{\bar
\rho}\Gamma_3)+{\tilde {\rm n}}_a\Gamma_1\right]~~.
\ee
Also, the spin connection evaluated on the background solution is:
\be
\partial_aX^M \omega_M{}^{AB}\Gamma_{AB}=2{\hat {\rm n}}_a
(\sinh{\bar \rho}\Gamma_0+\cosh{\bar \rho}\Gamma_3)\Gamma_1
\ee

The ${\bar\rho}$ dependence may be removed by a rotation (boost) in
the $(03)$ plane:
\be
\label{rotation}
&&S=\cosh\frac{\bar \rho}{2}+\sinh\frac{\bar \rho}{2}\Gamma_{03}
\\
&&(\cosh{\bar \rho}\Gamma_0+\sinh{\bar \rho}\Gamma_3)
=S\Gamma_0 S^{-1}
\cr
&&(\sinh{\bar \rho}\Gamma_0+\cosh{\bar \rho}\Gamma_3)
=S\Gamma_3 S^{-1}~~.
\nonumber
\ee
This is absorbed by a field redefinition of the fermions
\be
\theta^I=S\psi^I
\ee
which in turn introduces an additional connection component:
\be
\label{extra_conn}
S^{-1}\partial_aS=\frac{1}{2}{\tilde {\rm n}}_a\Gamma_{03}
\ee

Thus, we need to expand:
\be
{\cal L}^{IJ}_{ab}&=&{\bar\theta}^Ie\llap/{}_a 
\left(\partial_b+\frac{1}{4}
\omega_b{}^{AB}\Gamma_{AB}\right)\theta^J
+\frac{1}{8}e^{\phi}{\bar\theta}^Ie\llap/{}_a\left[F\llap/{}_{(2)}(i\sigma_2)^{JK}
+F\llap/{}_{(4)}(\sigma_1)^{JK}\right] e\llap/{}_b\theta^K
\cr
&=&\frac{R_{\rm string}}{2}\left[{\bar\psi}^I({\hat {\rm n}}_a\Gamma_0+{\tilde {\rm n}}_a\Gamma_1)
\left(\partial_b+\frac{1}{2}({\tilde {\rm n}}_b\Gamma_0-{\hat {\rm n}}_b\Gamma_1)\Gamma_3
\right)\psi^J\right.\cr
&+&\left.\frac{R_{\rm string}}{16}e^{\phi}{\bar\psi}^I\left[F\llap/{}_{(2)}(i\sigma_2)^{JK}
-F\llap/{}_{(4)}(\sigma_1)^{JK}\right] ({\hat {\rm n}}_a\Gamma_0+{\tilde
n}_a\Gamma_1)
({\hat {\rm n}}_b\Gamma_0+{\tilde {\rm n}}_b\Gamma_1)\psi^K\right]~~.
\label{non_summed_L}
\ee
In the flux term we used the fact that $F_{2}$ does not have
components in the AdS direction so it commutes with $\Gamma_0$ and
$\Gamma_1$ while $F\llap/{}_{(4)}\propto\Gamma_{0123}$ so it
anticommutes with $({\hat {\rm n}}_a\Gamma_0+{\tilde {\rm
n}}_a\Gamma_1)$. In the second term in the parenthesis, all factors of
$R$ and $k$ cancel out once the expressions of the dilaton and forms
are included.

Using the fact that $(\eta^{ab}\delta^{IJ}-\epsilon^{ab}s^{IJ})$ is
diagonal in the indices $I,J$ it is possible to simplify somewhat the
first term above, which we will denote by ${\cal D}_{ab}^{IJ}$. 
Indeed, opening the parenthesis,
\be
{\cal D}^{IJ}_{ab}&=&{\bar\psi}^I({\hat {\rm n}}_a\Gamma_0+{\tilde {\rm n}}_a\Gamma_1)
\left(\partial_b+\frac{1}{2}({\tilde {\rm n}}_b\Gamma_0-{\hat {\rm n}}_b\Gamma_1)\Gamma_3
\right)\psi^J\\
&=&{\bar\psi}^I({\hat {\rm n}}_a\Gamma_0+{\tilde {\rm n}}_a\Gamma_1)
\partial_b\psi^J-\frac{1}{2}({\hat {\rm n}}_a{\hat {\rm n}}_b+{\tilde {\rm n}}_a{\tilde
n}_b){\bar\psi}^I\Gamma_{013}\psi^J
+{\cal O}({\bar\psi}^I\Gamma_{3}\psi^J)
\nonumber
\ee
it is not hard to identify terms which vanish, if $I=J$, due to the
chirality of fermions.

The two terms arising in the sum the indices $I,J$ in 
$(\eta^{ab}\delta^{IJ}-\epsilon^{ab}s^{IJ}){\cal D}^{IJ}_{ab}$ are
both of the same type:
\be
(\eta^{ab}+\eta\epsilon^{ab}){\cal D}^{II}_{ab}=
-{\bar\psi}^I\Gamma_0(1-\eta\Gamma_0\Gamma_1)
\partial_0\psi^I+{\bar\psi}^I\Gamma_1(1-\eta\Gamma_0\Gamma_1)
\partial_1\psi^I
\ee
where we used the Virasoro constraint ${\hat {\rm n}}\cdot {\hat {\rm n}}+{\tilde {\rm n}}\cdot{\tilde
n}=0$. Here $\eta=-1$ if $I=1$ and $\eta=+1$ if $I=2$. It is useful to
note the explicit appearance of projection operators
\be
{\cal P}_\pm=\frac{1}{2}(1\pm\Gamma_{01})~~;
\label{proj_J0}
\ee
this is a consequence of the $\kappa$-symmetry of the action.

The trivial multiplication of vielbeine $e\llap/{}_ae\llap/{}_b=
({\hat {\rm n}}_a\Gamma_0+{\tilde {\rm n}}_a\Gamma_1) ({\hat {\rm n}}_b\Gamma_0+{\tilde {\rm n}}_b\Gamma_1)$
leads to a simple expression for the vielbein-dependent factor in the
flux-dependent term in (\ref{non_summed_L}). 
It is again a sum of two terms of the type
\be
&&(\eta^{ab}+\eta\epsilon^{ab})e\llap/{}_ae\llap/{}_b=2(1+\eta\Gamma_{01})
\ee
where 
we made use of the explicit expressions of the vectors $n$ and
${\tilde {\rm n}}$ to write 
$\epsilon^{ab}{\hat {\rm n}}_a{\tilde {\rm n}}_b=1$ and $-{\hat {\rm n}}\cdot
{\hat {\rm n}}+{\tilde {\rm n}}\cdot{\tilde {\rm n}}=+2$ and, as before, $\eta=-1$ for $I=1$ and
$\eta=+1$ for $I=2$. Note again the appearance of the projectors ${\cal
P}_\eta$. 

The action is easy to construct by starting from (\ref{lag_0});
skipping trivial details, the result is
\be
\frac{2}{iR_{\rm string}}L_{2F}&=&\frac{2}{R_{\rm string}}
(\eta^{ab}\delta^{IJ}-s^{IJ}\epsilon^{ab}){\cal L}^{IJ}_{ab}
\cr
&=& -{\bar\psi}^1\Gamma_0(1+\Gamma_0\Gamma_1) \partial_0\psi^1
+{\bar\psi}^1\Gamma_1(1+\Gamma_0\Gamma_1) \partial_1\psi^1\cr
&&
 -{\bar\psi}^2\Gamma_0(1-\Gamma_0\Gamma_1) \partial_0\psi^2
+{\bar\psi}^2\Gamma_1(1-\Gamma_0\Gamma_1) \partial_1\psi^2\cr
&+&
\frac{R_{\rm string}}{8}e^{\phi}{\bar\psi}^1\left[F\llap/{}_{(2)}(+1)
-F\llap/{}_{(4)}(+1)\right] (1-\Gamma_{01})\psi^2\cr
&+&
\frac{R_{\rm string}}{8}e^{\phi}{\bar\psi}^2\left[F\llap/{}_{(2)}(-1)
-F\llap/{}_{(4)}(+1)\right] (1+\Gamma_{01})\psi^1~~.
\ee

At this stage it is useful to recall that $\psi^1$ and $\psi^2$ are
spinors of opposite chirality -- with $\Gamma_{-1}$ the ten-dimensional chirality operator,
  $\Gamma_{-1}\psi^1=\psi^1$ and
$\Gamma_{-1}\psi^2=-\psi^2$  -- and thus may be assembled into a single,
non-chiral ten-dimensional spinor 
$\psi=\psi^1+\psi^2$. In terms of this new field the action takes a
very simple form:
\be
L_{2F}=\frac{iR_{\rm string}}{2}
\left(2{\bar\psi}(-\Gamma^0\partial_0 +
\Gamma^1\partial_1){\cal P}_+\psi
-\frac{R_{\rm string}}{4}e^\phi{\bar\psi} \left[F\llap/{}_{(2)}\Gamma_{-1}
+F\llap/{}_{(4)}\right]{\cal P}_+\psi\right)~~.
\label{final_S0}
\ee

This action is still invariant under local $\kappa$-transformations, a 
fact reflected by the manifest appearance of a projector ${\cal
P}_+$ in all terms in the action. It is only natural to choose the gauge
\be
{\cal P}_+\psi=\psi~~,
\ee
which eliminates from the fermion fields the components not appearing
in the Lagrangian. This algebraic gauge, which is similar to the
light-cone gauge, introduces no $\kappa$-symmetry ghosts.

For explicit calculations it is necessary to expand also the last term
in the action (\ref{final_S0}) using the explicit form of the form
fields; the relative factor of $R_{\rm string}$ with the derivative
term cancels out and we find
\be
-\frac{R_{\rm string}}{4}\;e^\phi{\bar\psi} \left[F\llap/{}_{(2)}\Gamma_{-1}
+F\llap/{}_{(4)}\right]\psi =
-\frac{1}{4}\left[+2(\Gamma_{45}-\Gamma_{67}+\Gamma_{89})\Gamma_{-1}
+6\Gamma_{0123}\right]~~.
\label{flux}
\ee

The spectrum of fermion quadratic operator (\ref{final_S0}) may be
found by evaluating its eigenvalues and setting them to zero. It turns
out that there are two massless and six massive modes with unit mass:
\be
\omega_{1,2}(n)=|n|~~~~~~~~\omega_{3, 4, 5,6,7,8}(n)=\sqrt{n^2+\kappa^2}~~.
\ee
Note that, similarly to the bosonic spectrum, $\sum_{i=1}^8m_i^2=6\kappa^2$;
therefore, the one-loop correction to the energy of the $(S,J=0)$
string is finite. We will evaluate it in section (\ref{sec:energies}).

The structure of this spectrum could have been anticipated from
symmetry considerations. Indeed, as reviewed in the introduction, the
supersymmetries form a ${\mathbf 6}_0\oplus {\mathbf 1}_2
\oplus {\mathbf 1}_{-2}$ representation of the global symmetry group 
$SU(4)\times U(1)$. Thus, we should expect six modes of equal
masses. An additional $\IZ_2$ (charge conjugation) symmetry changing
the sign of the $U(1)$ charges suggests that the remaining two modes
should also have equal masses.

\subsubsection{The $(S,J\neq 0)$ string}

The inclusion of a single angular momentum on $\IP^3$ is technically
quite straightforward. The main difference is that now all three
vectors (\ref{vectors}) are nontrivial and given by 
${\hat {\rm n}}=(\kappa,0)$, ${\tilde {\rm n}}=(0,\mu )$ 
and ${\hat {\rm m}}=(\nu,0)$. 

Since the angular momentum on $\IP^3$ is described by a linear profile
along an isometry direction, it introduces no additional worldsheet
coordinate dependence in the fermion action besides the one due to the
AdS$_4$ part of the solution. As for ${J}=0$ this latter dependence
may be eliminated by the rotation (\ref{rotation}). After this
rotation, the vielbein and the spin connection modified to include
the effects of the rotation (\ref{extra_conn}) 
are:
\be
e\llap/{}_a=\frac{R_{\rm string}}{2}\left[
{\hat {\rm n}}_a\Gamma_0+{\tilde {\rm n}}_a\Gamma_1+{\hat {\rm m}}_a\Gamma_9\right]
~~~~~~~~
\frac{1}{4}{\widetilde\omega}_a{}^{AB}\Gamma_{AB}=
\frac{1}{2}\left({\tilde {\rm n}}_a\Gamma_0 - {\hat {\rm n}}_a\Gamma_1\right)\Gamma_3
\ee
\be
\frac{1}{4}e\llap/{}_a{\widetilde\omega}_b{}^{AB}\Gamma_{AB}
=\frac{R_{\rm string}}{2}\left[
-\frac{1}{2}({\hat {\rm n}}_a{\hat {\rm n}}_b+{\tilde {\rm n}}_a{\tilde {\rm n}}_b)\Gamma_{013}
+\frac{1}{2}{\hat {\rm m}}_a{\tilde {\rm n}}_b \Gamma_{039}
-\frac{1}{2}{\hat {\rm m}}_a{n}_b \Gamma_{139}\right]~~.
\ee

The two terms arising from the gravitational covariant derivative
continue to have a similar structure, up to some signs (denoted by
$\eta$) which again related to the chirality of
the spinors:
\be
\label{derterms}
\frac{2}{R_{\rm string}}(\eta^{ab}+\eta\epsilon^{ab}){\cal D}^{II}_{ab}
&=&(\eta^{ab}+\eta\epsilon^{ab}){\bar\psi}^I({\hat {\rm n}}_a\Gamma_0+{\tilde
n}_a\Gamma_1 +{\hat {\rm m}}_a\Gamma_9)
\partial_b\psi^I \\
&-&
 \frac{1}{2}({\hat {\rm n}}\cdot {\hat {\rm n}}+{\tilde {\rm n}}\cdot{\tilde {\rm n}})
\;{\bar\psi}^I\Gamma_{013}\psi^I
-\frac{1}{2}{\hat {\rm m}}\cdot {\hat {\rm n}}\;{\bar\psi}^I\Gamma_{139}\psi^I
+\frac{\eta}{2} {\hat {\rm m}}\times {\tilde {\rm n}}\;{\bar\psi}^I\Gamma_{039}\psi^I
\nonumber
\\
&=&-{\bar\psi}^I(\kappa\Gamma_0+\eta\mu \Gamma_1 +\nu\Gamma_9)
\partial_0\psi^I
        +{\bar\psi}^I(\eta\kappa\Gamma_0+\mu \Gamma_1 +\eta\nu\Gamma_9)
\partial_1\psi^I\\
&+&
 \frac{1}{2}\nu^2\;{\bar\psi}^I\Gamma_{013}\psi^I
+\frac{1}{2}\kappa\nu\;{\bar\psi}^I\Gamma_{139}\psi^I
+\frac{\eta}{2} \nu\mu \;{\bar\psi}^I\Gamma_{039}\psi^I~~.
\nonumber
\ee
It is easy to identify in the derivative terms a 
projector (${\cal P}_\eta^2={\cal P}_\eta$) analogous to the one in
equation (\ref{proj_J0}); it is:
\be
{\cal P}_\eta=
\frac{1}{2}\left(1+\eta\left(\frac{\kappa}{\mu }\Gamma_0
+\frac{\nu}{\mu }\Gamma_9\right)\Gamma_1\right)~~~~\eta=\pm~~.
\ee
Using it and introducing the same unconstrained, non-chiral
ten-dimensional spinor as before 
$\psi=\psi^1+\psi^2$
the equation (\ref{derterms}) can be reorganized as:
\be
(\eta^{ab}\delta^{IJ}-\epsilon^{ab}s^{IJ}){\cal D}^{IJ}_{ab}&=&
\frac{R_{\rm string}}{2}{\bar\psi}\left[
-2(\kappa\Gamma_0+\nu\Gamma_9)\,{\cal P}_+\,\partial_0+2\mu \Gamma_1\,
{\cal P}_+\,\partial_1\right.\cr
&&~~~~~~~~~~~
\left.+\frac{1}{2}\nu\left(\nu\Gamma_0
+\kappa\Gamma_9\right)\Gamma_{13}
-\frac{1}{2}\nu\mu \Gamma_{039}\right]\psi~~.
\label{der_J}
\ee
Note that, unlike the string spinning only in AdS, there is a
nontrivial connection term; these terms vanish as $\nu\sim
J\rightarrow 0$ and the derivative terms reduce to those of the 
previous section.

To simplify the flux contribution it is useful to use the explicit 
forms of the vectors $n,\,{\tilde {\rm n}}$ and $m$ and to split the
2-form into a part depending on the $\IP^3$ isometry direction
(i.e. $\Gamma_9$), $F\llap/{}_2^{(1)}$, and the rest,
$F\llap/{}_2^{(2)}$:
\be
F\llap/{}_2=F\llap/{}_2^{(1)}+F\llap/{}_2^{(2)}~~.
\ee
In terms of these components, the flux terms are:
\be
&&\left(\frac{2}{R_{\rm string}}\right)^2e^\phi (\eta^{ab}\delta^{IJ}-\epsilon^{ab}\epsilon^{IJ}){\bar\psi}^Ie\llap/{}_a\left[F\llap/{}_{(2)}(i\sigma_2)^{JK}
+F\llap/{}_{(4)}(\sigma_1)^{JK}\right] e\llap/{}_b\psi^K\cr
&&~~~~~~~~~~~~~~~~~~~~~~~~~~~
= {\bar \psi}F\llap/{}_2^{(2)}\Gamma_{-1}\left(2\mu ^2 +2\kappa\mu  \Gamma_{01}\Gamma_{-1} -2\mu \nu\Gamma_{19}\Gamma_{-1} \right)\psi\cr
&&~~~~~~~~~~~~~~~~~~~~~~~~~~~
+{\bar \psi}(F\llap/{}_2^{(2)}\Gamma_{-1}-F\llap/{}_4)\left(2\kappa^2 -2\kappa\nu\Gamma_{09} +2\kappa\mu  \Gamma_{01}\Gamma_{-1}\right)\psi
\ee

It is not hard to expose the projectors in this expression; restoring
the numerical coefficient of the flux term in the covariant derivative
and making use of the explicit expressions for the form fields we find
that the contribution of the form fields to the fermion action to
quadratic order in fermions and to leading order in the expansion in
bosonic fluctuations is
\be
&&
\frac{1}{8}e^\phi (\eta^{ab}\delta^{IJ}-\epsilon^{ab}\epsilon^{IJ}){\bar\psi}^Ie\llap/{}_a\left[F\llap/{}_{(2)}(i\sigma_2)^{JK}
+F\llap/{}_{(4)}(\sigma_1)^{JK}\right] e\llap/{}_b\psi^K\cr
&=&\frac{R_{\rm string}}{2}\left[
\frac{1}{16} (4\mu ^2){\bar\psi}(-\Gamma_{57}+\Gamma_{68})\Gamma_{-1}{\cal P}_+\psi
  +\frac{1}{16}
(4\mu \kappa){\bar\psi}(-\Gamma_{49}\Gamma_{-1}+3\Gamma_{0123}){\cal
P}_+\psi
\right]~~.
\label{flux_J}
\ee

Combining the derivative (\ref{der_J}) and the flux terms
(\ref{flux_J}) it is easy to find the relevant gauge-invariant
Lagrangian (a constant rotation in the $(09)$ plane may be used to
slightly simplify the derivative term):
\be
\frac{2}{iR_{\rm string}}{\cal L}_{2F}&=&{\bar\psi}\left[
-2(\kappa\Gamma_0+\nu\Gamma_9)\,{\cal P}_+\,\partial_0+2\mu \Gamma_1\,
{\cal P}_+\,\partial_1
+\frac{1}{2}\nu\left(\nu\Gamma_0
+\kappa\Gamma_9\right)\Gamma_{13}-\frac{1}{2}\kappa\mu \Gamma_{039}\right]\psi
\cr
&-&\frac{1}{16}\left[(4\mu ^2){\bar\psi}(-2\Gamma_{57}+2\Gamma_{68})\Gamma_{-1}{\cal P}_+\psi
  +
(4\mu \kappa){\bar\psi}(-2\Gamma_{49}\Gamma_{-1}+6\Gamma_{0123}){\cal
P}_+\psi
\right]~~.\nonumber\\
\ee
As before, the manifest appearance of ${\cal P}_+$ suggests that a
natural gauge condition is
\be
{\cal P}_+\psi=\psi~~,
\ee
in analogy to the $J=0$ analysis. As in that case, this gauge
condition does not introduce any $\kappa$-symmetry ghosts. It is
moreover easy to see that the limit $\nu\rightarrow 0$ quickly leads
to the equation (\ref{final_S0}), implying that the gauge condition is
a smooth function of $J$.

The energy spectrum of quadratic fluctuations can be found by first
setting to zero the eigenvalues of the quadratic fluctuations operator;
the result, which may be checked by a variety of means, is that
\be
\label{eqn:fermionic_freq}
\omega_{1,2,\pm}(n)&=&\pm\frac{\nu}{2}+\sqrt{n^2+\kappa^2}
\cr
\omega_{3,4,\pm}(n)&=&\frac{1}{\sqrt{2}}
\sqrt{\kappa^2+2n^2\pm\sqrt{\kappa^4+4\nu^2 n^2}}
\ee
Thus, we find four modes with unit mass and the other four modes have
more complicated dispersion relations which are similar to those for
some of the bosonic AdS fluctuations (\ref{eqn:bosonic_ads_freq}). 

It is interesting to note that the massless fermion modes present at
$J=0$ are now lifted. The fact that four modes continue to have equal
masses (up to a time-dependent rotation of their wave
functions) is consistent with (and in fact should be expected from)
the fact that the worldsheet background breaks the 
symmetry of $\IP^3$ from $SO(6)$  to $SO(4)$. 

\section{One-loop correction to string energies}
\label{sec:energies}

Given the spectrum of fluctuations we found in previous sections, the
one-loop correction to the string energy may be computed in a variety
of ways. An important subtlety is that the relation between the
parameters of the solution and the field theory charges may receive
quantum corrections. Such effects may be captured either in the
Hamiltonian formalism \cite{Frolov:2002av} or in the Lagrangian
formalism \cite{Roiban:2007ju}. In the latter approach the fundamental
quantity is the worldsheet partition function in the presence of
chemical potentials for all charges. The target space energy is found
by Legendre-transforming the logarithm of the partition function with
respect to the chemical potentials. In the process one also uses the
quantum Virasoro constraint, which sets to zero the quantum expectation
value of the worldsheet Hamiltonian.

The results obtained through these two methods imply that such
modifications to the relation between charges and parameters of the
classical solution are irrelevant in a one-loop calculation. It is
perhaps more convenient to use the expression for the string energy in
conformal gauge in terms of the fluctuation fields derived in Appendix
A of \cite{Frolov:2002av}:
\be
E_1=\frac{1}{\kappa}\langle\Psi|H_2|\Psi\rangle
\ee
with $H_2=\int \tfrac{d\sigma}{2\pi} {\cal H}_2({\tilde t},{\tilde
\phi},\dots)$ being the quadratic worldsheet Hamiltonian corresponding
the fluctuation action at this order. For the spinning string the
classical solution spontaneously breaks supersymmetry and we expect to
find a non-trivial correction at one-loop. We begin with the simpler
$(S,J=0)$ case and then proceed to the general solution.

\subsection{$(S,J=0)$}

  For the case $(S,J=0)$ we
have in the scaling limit that the energy is given by the sum over frequencies
\be
E_1=\frac{1}{2\kappa}\sum_{n=-\infty}^{\infty}K_n 
\ee 
with $K_n=\sqrt{n^2+4 \kappa^2}+\sqrt{n^2+2 \kappa^2}+6 \sqrt{n^2}-6
\sqrt{n^2+ \kappa^2}-2 \sqrt{n^2}$. In the scaling limit 
($\kappa\gg1$) this sum can be replaced by an integral. After
rescaling the worldsheet mode numbers, $n$, and introducing the
continuous worldsheet momentum, $p$, we have
\be
E_1=\kappa \int_{0}^{\infty}dp\ \sqrt{p^2+4}+\sqrt{p^2+2}+6 \sqrt{p^2} -6 \sqrt{p^2+1} -2\sqrt{p^2}+{\cal O}\left(\kappa^0\right).
\ee
It is straightforward to evaluate this integral by imposing a cutoff,
performing the individual integrals and taking the cutoff to infinity.
Expanding at large values of the cutoff one can check that the
quadratic and logarithmic UV divergences vanish. The leading finite
piece is given by
\be
\label{cusp_CS_1}
E_1&=&-\kappa\ \frac{5}{2} \ln 2+{\cal O}\left(\kappa^0\right)\nn\\
& =&-\frac{5\ln 2}{2\pi}\ln S+ {\cal O}\left(\ln^0 S\right)~~.
\ee

Thus we see that, as for the AdS$_5$ case, the one-loop piece
continues to scale as $\ln S$ and there is no stronger $\ln^\alpha
S$, $\alpha> 1$, dependence. In fact we expect, not least on simple
dimensional grounds, that this structure will continue to all orders
at strong coupling and can be interpolated to match the weak coupling
result.

\subsection{$(S,J\neq0)$}
We can now use essentially the same method for the generalized $(S,J)$
string solution with two non-vanishing charges. In this case the sum
of frequencies of the bosonic fluctuations,
\eqref{eqn:bosonic_ads_freq} and below, and the fermionic
fluctuations, \eqref{eqn:fermionic_freq}, is
\be
K_n &=& \sqrt{n^2+2\kappa^2+2 \sqrt{\kappa^4+ n^2 \nu^2}}
+\sqrt{n^2+2\kappa^2-2\sqrt{\kappa^4 + n^2 \nu^2}}\nn\\
& &
+\sqrt{n^2+2\kappa^2-\nu^2}+4\sqrt{n^2+\frac{\nu^2}{4}}
+\sqrt{n^2+\nu^2}-4\sqrt{n^2+\kappa^2}\nn\\
& &-2\left(
  \frac{1}{\sqrt{2}} \sqrt{2 n^2+\kappa^2+\sqrt{\kappa^4 +4 n^2 \nu^2}}
+ \frac{1}{\sqrt{2}} \sqrt{2 n^2+\kappa^2-\sqrt{\kappa^4 +4 n^2
\nu^2}}
    \right).
\ee
We again replace the discrete sum over mode numbers by an integral
which, with the help of identities and changes of variables from
appendix \ref{app:integrals}, results in a one-loop correction to the
energy of
\be
E_1&=&\frac{\nu}{2u}\Big[-(1-u^2)+\sqrt{1-u^2}-2u^2\ln u\nn\\
& & \kern+40pt-(2-u^2)\ln \left(\sqrt{2-u^2}(1+\sqrt{1-u^2})\right)
-2(1-u^2) \ln 2\Big]~~.
\label{cusp_CS_1_with_J}
\ee
which is seen to be remarkably similar to the \AdSS  result
though with some modifications. Here we have used the parameter
\be
u=\frac{\nu}{\kappa}=\frac{\ell}{\sqrt{1+\ell^2}},\qquad 
\ell=\frac{\nu}{\mu} =\frac{{\cal J}}{2 \ln {\cal S}}
\ee
and it is straightforward to see that in the $u\rightarrow0$ limit it
reduces to equation (\ref{cusp_CS_1}). This generalized scaling function is a
useful tool in studying the AdS$_5$/CFT$_4$ duality and it is to be
expected that it will also be so in the case at hand.

\section{Comparison with the Bethe Ansatz and outlook}

The dilatation operator of the ${\cal N}=6$ Chern-Simons theory was shown
\cite{Minahan:2008hf, Bak:2008cp}, 
to leading order in the scalar sector, to be equivalent to the
Hamiltonian of an integrable (alternating) spin chain. It was moreover
argued that the worldsheet theory in the dual supergravity background
is also classically integrable \cite{Arutyunov:2008if,
Stefanski:2008ik}. It is tempting to infer that integrability
potentially exists for finite values of the 't~Hooft coupling as
well. This conjecture is based on the nontrivial assumption that the
anomaly of the conservation of the hidden charges present in the
bosonic $\IP^3$ sigma model is canceled in the full Green-Schwarz
theory. It would be important to have direct tests of this assumption.

With this starting point, and using the observation that the
transformation rules of the spin chain excitations are similar to
those of the spin chain excitations in ${\cal N}=4$ SYM, all-order
Bethe equations have been conjectured \cite{Gromov:2008qe} for the
${\cal N}=6$ Chern-Simons theory. As in ${\cal N}=4$ SYM, the tensor
structure of the relevant scattering matrices is fixed by
symmetries. The difference compared to the four-dimensional case is
that the magnon dispersion relation acquires an overall numerical
factor and in both the magnon dispersion relation and the S-matrix 
the 't~Hooft coupling enters through an arbitrary function
$h(\lambda)$\footnote{This function may be fixed by a direct calculation
of the magnon dispersion relation in the ${\cal N}=6$ Chern-Simons theory.}
\be
\epsilon(p)=\frac{1}{2}\sqrt{1+16h(\lambda)^2\sin^2\frac{p}{2}}~~.
\ee
In ${\cal N}=4$ SYM one has $h(\lambda)=\sqrt{\lambda}/{4\pi}$ while in
the ${\cal N}=6$ Chern-Simons theory 
\be
h(\lambda)=\left\{
\begin{array}{ll}
\lambda+{\cal O}(\lambda^3)&~~\lambda\ll1\cr
\sqrt{\frac{\lambda}{2}}+{\cal O}(1)&~~\lambda\gg 1
\end{array}
\right.~~.
\ee
It was further argued
that, up to the same function $h(\lambda)$, the dressing phase is the
same as that of the scattering matrix of the ${\cal N}=4$ SYM spin
chain. 

This relation between scattering matrices and dispersion relations
implies in turn that most anomalous dimensions in the ${\cal N}=6$
Chern-Simons theory enjoy simple relations with those of ${\cal N}=4$
SYM theory. For example, it was argued in \cite{Gromov:2008qe} that
for the universal scaling functions this relation is
\be
f_{CS}(\lambda)=\frac{1}{2}f_{{\cal N}=4}(\lambda)
\Big|_{\sqrt{\lambda}\mapsto 4\pi h(\lambda)}~~.
\ee
Using the result from the algebraic curve calculation 
\cite{Shenderovich:2008bs} that the constant
term in $h(\lambda)$ vanishes
in the regularization scheme adapted to the algebraic curve
calculation
and of the known strong coupling expansion of the universal scaling
function,
\be
f_{{\cal N}=4}(\lambda)=\frac{1}{\pi}
\left(\sqrt{\lambda}-3\ln 2+{\cal O}
\left(\textstyle{\frac{1}{\sqrt{\lambda}}}\right)\right)~~,
\label{Neq4_cusp}
\ee
it is easy to find that 
\be
f_{CS}(\lambda)=\frac{1}{2}f_{{\cal
N}=4}(\lambda)\Big|_{
\sqrt{\lambda}\mapsto 4\pi h(\lambda)}=\sqrt{2\lambda}-\frac{3\ln
2}{2\pi} + {\cal O}
\left(\textstyle{\frac{1}{\sqrt{\lambda}}}\right)
\ee
The first term matches (by construction) the leading order in the
strong coupling expansion of the spinning folded string energy
(\ref{cusp_CS_0}). The second term above however departs from the
worldsheet predictions (\ref{cusp_CS_1}) for the next-to-leading 
order correction to the universal scaling function.

In the same spirit one may compare (\ref{cusp_CS_1_with_J}) with the
consequence of the conjectured Bethe ansatz for the ${\cal N}=6$
Chern-Simons theory. Instead of \footnote{We are grateful to P. Vieira
for pointing out a difference of 2 in our definition of $J$ and that
used in \cite{Gromov:2008qe}.}
\be
f_{CS}(\lambda,\frac{J}{\ln S})=\frac{1}{2}f_{{\cal
N}=4}(\lambda,\frac{J}{\ln S})\Big|_{
\sqrt{\lambda}\mapsto 4\pi h(\lambda)}~~,
\ee
it is easy to see that the leading and next-to-leading terms in the
string coupling expansion of the generalized scaling function
$f_{CS}(\lambda,\ell)$ are consistent with
\be
f_{CS}(\lambda,\frac{\cal J}{\ln {\cal S}})=\frac{1}{2}f_{{\cal
N}=4}(\lambda,\frac{\cal J}{\ln {\cal S}})\Big|_{
\sqrt{\lambda}\mapsto 4\pi h(\lambda)}-\frac{\nu}{u}(1-u^2) \ln 2
\ee
where $f_{{\cal N}=4}(\lambda,\frac{\cal J}{\ln {\cal S}})$ is given
in \cite{Frolov:2006qe,Frolov:2002av}.

Though the resolution of this puzzle is not immediately apparent,
several possibilities present themselves. For example, it may be possible
that twist-two operators dual to the spinning folded string have been
misidentified. 
It may also be possible that the problem lies either with the assumption
that integrability survives beyond the leading order in the strong
coupling expansion or with the precise expression for the scattering
matrix. Since its tensor structure is determined by symmetries whose
action is closely related to the action of symmetry generators in
${\cal N} = 4$ SYM, it may be that the dressing phase receives
additional next-to-leading order corrections. 
Perhaps a profitable route to finding these corrections is to follow
the strategy of \cite{Hernandez:2006tk} and construct the phase by
matching it with the one-loop corrections to the circular string
rotating entirely in AdS$_4$.
Another approach would, of course, be a direct solution of the crossing
equation. The similarity of the symmetry groups of the scattering
matrix of the worldsheet theory in \AdSP and \AdSS suggests however
that the correction to the dressing phase, if any, is a solution of
the homogeneous crossing equation.

For the spinning string in AdS$_5 \times$ S$^5$ it has proven possible
to extend the calculation of the quantum corrections to two-loops
\cite{Roiban:2007jf,Roiban:2007dq,Roiban:2007ju} and it would
certainly be interesting to repeat that calculation in the current
context. In the absence of an argument that there exists a
$\kappa$-gauge in which the action becomes quadratic in fermions, a
prerequisite for a higher-loop calculation is knowledge of the
contributions to the string action from terms quartic (and higher) in
fermions. One could hope to possibly use the $OSp(6|4)$ coset
sigma-model \cite{Arutyunov:2008if,Stefanski:2008ik,Fre:2008qc} though
due to the enhanced $\kappa$-symmetry at $J=0$, such a calculation 
appears challenging at first sight. Alternatively one can derive the type
IIA string action by doubly dimensionally reducing the supermembrane
action \cite{Duff:1987bx,Tseytlin:1996hs,Cvetic:1999zs}
\be
S=-\int d^3\zeta \sqrt{-{\rm det}\ g(Z(\zeta))}+\int_{{\rm M}_3} B
\ee
where $Z=(X^\mu,\theta^\alpha)$ are the eleven dimensional target
superspace coordinates, $\zeta=(\tau,\sigma, \sigma_3)$ are the
worldvolume coordinates,
\be
g_{{\hat i}{\hat j}}=\partial_{\hat i} Z^M\partial_{\hat j} 
Z^N E^{\hat r}_M E^{\hat s}_N \eta_{\hat{r}\hat{s}}
\ee
is the pullback of the supervielbein to the worldvolume and $B$ is the
pullback of the super-three-form. This procedure can be somewhat
involved and has been explicitly done only to quadratic order in
fermions for generic bosonic backgrounds. However for the case of
$AdS_4\times S^7/{\mathbf Z}_k$ due to the large degree of symmetry it
may be possible to carry it out to higher orders starting from the
supermembrane action of \cite{deWit:1998yu} where explicit expressions
for the supervielbein and $B$ are given to all orders in fermions.

A further appeal of such an approach relates to the exactness of the
AdS$_4\times $S$^7$ geometry and its consequences. As was argued by
Kallosh and Rajaraman \cite{Kallosh:1998qs} the $AdS_4\times S^7$
geometry is exact in that it cannot receive $\ell_{p}$ corrections
which are consistent with supersymmetry. 
While the $\IZ_k$ orbifold relating it to \AdSP breaks some of the
supersymmetry for $k>2$, it is reasonable to expect that this geometry
remains unchanged and thus that the type IIA solution $AdS_4\times
\IP^3$ does not receive $\alpha'$ corrections (up to perhaps a finite
renormalization of the radius of the space).

Another possible approach to extracting higher-loop information is
suggested by the work of Alday and Maldacena who showed,
\cite{Alday:2007mf}, that for AdS$_5\times $S$^5$ the leading
logarithmic dependence on $u$ is described by a two-dimensional
$O(6)$ sigma model. 
At the level of the string worldsheet one may justify this by
integrating out the massive modes and constructing in this way an
effective action for the light modes \footnote{``Light'' stands for
masses of order $\nu$ or $u$.}.  
Similar reasoning suggests that here the leading logarithmic
dependence in $u$ may be captured by a $\IP^3$ model coupled to two light
fermions - the light degrees of freedom in the current model. While it
is known that the bosonic $\IP^3$ model is not integrable at the
quantum level due to an anomaly in the conservation of the non-local
charges, \cite{Abdalla:1980jt}, it is possible to couple the theory to
fermions such that the anomaly cancels. Such are the
minimal or the supersymmetric couplings, see for example
\cite{Abdalla:1983ae}; it would be interesting to check whether the same is 
true in this case.
One would then be able to predict the coefficients of the leading and
first subleading $\ln u$ terms to all orders in the strong coupling
expansion.

\bigskip
\subsection*{Acknowledgments}
\bigskip
We would like to thank F.~Alday, G.~Arutyunov, N.~Beisert, S.~Frolov, 
M. Spradlin, A.~Tseytlin and B.~Zwiebel for discussions. We would also
like to thank N.~Gromov and P.~Vieira for useful comments on the
manuscript. The work of RR is supported by the Department of Energy
OJI award DE-FG02-90ER40577, the National Science Foundation under
grant PHY-0608114 and the A. P. Sloan Foundation.


\appendix


\section{Integrals}
 \label{app:integrals}

The sum of frequencies of bosonic and fermionic fluctions can be put
in a form which can be integrated without difficulty using the
following identities \cite{Frolov:2006qe}:
\be
S_B(p)&=&\sqrt{2+p^2+2\sqrt{1+p^2u^2}}+\sqrt{2+p^2-2\sqrt{1+p^2u^2}}\cr
&=&\sqrt{4u^2+(p+\sqrt{p^2+4(1-u^2)})^2}\\[5pt]
S_F(p)&=&\frac{1}{\sqrt{2}}\sqrt{1+2p^2+\sqrt{1+4p^2u^2}}
             +    \frac{1}{\sqrt{2}}\sqrt{1+2p^2-\sqrt{1+4p^2u^2}}\cr
&=&\sqrt{u^2+(p+\sqrt{p^2+(1-u^2)})^2}~~.
\ee
Using a cutoff regularization for the integral over $p$ and changing
the integration variable $z=p+\sqrt{p^2+4(1-u^2)}$ and
$z=p+\sqrt{p^2+(1-u^2)}$, respectively, the integrals become
\be
&&\int_0^L dp S_B(p) = \int_{\sqrt{4(1-u^2)}}^{L+\sqrt{L^2+4-4u^2}}
\frac{dz}{z}\left(\frac{4-4u^2}{z}+z\right)
\sqrt{4u^2+z^2}
\cr
&&\int_0^L dp S_F(p) = \int_{\sqrt{1-u^2}}^{L+\sqrt{L^2+1-u^2}}  
\frac{dz}{z}\left(\frac{1-u^2}{z}+z\right)\sqrt{u^2+z^2}
\ee
which can be straightforwardly evaluated.

\vfill
\newpage

\bibliographystyle{nb}
\bibliography{String_Action}

\end{document}